# SECURING A QUANTUM KEY DISTRIBUTION RELAY NETWORK USING SECRET SHARING

*S.M. Barnett[1] & S.J.D. Phoenix[2]*

[1]Dept of Physics, Strathclyde University, Glasgow, UK
[2]Khalifa University, Abu Dhabi Campus, PO Box 127788, Abu Dhabi, UAE

**ABSTRACT**

We present a simple new technique to secure quantum key distribution relay networks using secret sharing. Previous techniques have relied on creating distinct physical paths in order to create the shares. We show, however, how this can be achieved on a single physical path by creating distinct logical channels. The technique utilizes a random 'drop-out' scheme to ensure that an attacker must compromise *all* of the relays on the channel in order to access the key.

## 1. INTRODUCTION

From its beginnings as a theoretical curiosity some two decades ago Quantum Key Distribution (QKD) is now a commercially available technology that is currently undergoing trials in a number of locations [1]. The technology offers a way of establishing a random sequence of binary digits between two end users in such a way that the secrecy of the established bit strings can be guaranteed. This bit string can then be used as a key in cryptographic applications.

One of the biggest obstacles to the widespread introduction of QKD techniques is the distance limitation in optical fibre which restricts current applications to a few tens of kilometers. The distance can be extended by using relays. In standard commercial applications these relays establish link-by link keys. We have shown elsewhere how relays can be adapted to provide end-to-end key distribution [2].

The major difficulty with relays from a security perspective is that they must be trusted; compromise of one relay will compromise the entire channel. Secret sharing schemes have been proposed to overcome this [3], but they rely on a network topology that admits the creation of distinct physical paths upon which the shares can be transmitted. We show here how a relay-based QKD network can be secured such that an attacker has to compromise *all* of the relays on the channel in order to access the information about the key. The basic principle is to create redundancy on the channel by using more relays than are strictly necessary for overcoming the distance limitation. This redundancy can then be utilized to create the distinct logical channels necessary for implementing a secret sharing technique.

## 2. THE BASIC SCHEME

In order to illustrate the technique we consider the channel shown in figure 1 consisting of the transmitter (Alice), the receiver (Bob), and 3 intermediate relays $R_j$ such that only one of the relays is actually needed to provide the requisite distance extension. We consider here the pass-through relays of [2] but the method is also applicable to relays when operated in link-by-link mode.

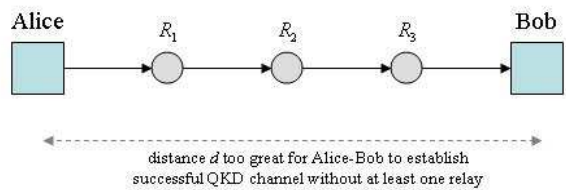

**Figure 1.** A QKD channel between Alice and Bob with 3 relays. Alice and Bob need at least one of these relays in order to successfully exchange a quantum key.

The channel in Figure 1, if all of the relays are involved in the communication, is vulnerable to the compromise of just one of the relays. All of the relays, in this case, have to be trusted devices. One way to mitigate this is to drop out the relays at random for each time slot. By drop out here we mean that the relay is switched off in such a way as to be completely transparent to the channel. Thus in any given time slot we may have 1, 2, 3, or none of the relays operating.

It is straightforward to see that compromise of only one relay in this instance reduces the information available to the eavesdropper. With compromise of one relay the eavesdropper is able to access 4/7 of the information about the key for the channel of Figure 1. This is still too much to allow a successful quantum key distribution.

The facility to drop out the relays at random, however, allows a more sophisticated approach. In Table 1 we list the possible logical channels created by the simple expedient of having the relays on or off at random. Each of these channels represents a unique quantum key distribution. Only Alice and Bob have access to *all* of these logical channels. In the last column of Table 1 we write the quantum key established using an obvious notation.

**Table 1.** The possible logical channels created using the random drop-out of relays. Note that when all relays are 'off' it is not possible for Alice and Bob to exchange a key because the distance is too great to successfully use a QKD technique. The open squares indicate when a relay is in the 'off' position.

| Alice | $R_1$ | $R_2$ | $R_3$ | Bob | Key |
|---|---|---|---|---|---|
| ■ | □ | □ | □ | ■ | — |
| ■ | ■ | □ | □ | ■ | $K_1$ |
| ■ | □ | ■ | □ | ■ | $K_2$ |
| ■ | □ | □ | ■ | ■ | $K_3$ |
| ■ | ■ | ■ | □ | ■ | $K_{12}$ |
| ■ | ■ | □ | ■ | ■ | $K_{13}$ |
| ■ | □ | ■ | ■ | ■ | $K_{23}$ |
| ■ | ■ | ■ | ■ | ■ | $K_{123}$ |

Alice and Bob now consider each of these keys as shares in a secret sharing scheme [4] and form a final key by taking the binary addition of these. The final quantum key is therefore

$$K = K_1 \oplus K_2 \oplus K_3 \oplus K_{12} \oplus K_{13} \oplus K_{23} \oplus K_{123} \quad (1)$$

None of the relays alone is able to construct this key. Indeed, no two relays can construct the key either. An attacker would have to compromise *all* of the relays in the channel in order to be able to obtain the key.

The situation is even more problematical for an eavesdropper. Let us suppose that she has compromised one of the relays and tries to make it operate for every timeslot. In other words, she disables the random drop-out feature and tries to fool the system that the relay is actually dropping out as expected. Because of the quantum nature of the transmission this behaviour can be detected in the same way the presence of an eavesdropper can be detected on a standard QKD channel. The eavesdropper has no choice but to randomly turn off the relay under her control so that the statistics match those expected. Any deviation can be readily detected. For a given relay, using this simple scheme, we should expect to find it on for about half of the timeslots on average.

It is also possible to consider more complicated drop-out statistics where the frequency of the devices being on or off can be adjusted. This can also be used to detect the presence of an active eavesdropper. As in the case of normal QKD a passive eavesdropper has no chance of ever obtaining the eventual key.

If the presence of an eavesdropper is suspected then Alice and Bob may choose to suspend the transmission. However, if the analysis shows that one of the relays is compromised then Alice and Bob can choose channels in which this relay does not play a part (although this is not strictly necessary because of the properties of the secret sharing).

Let us consider now the channel shown in Figure 1 when at least 2 relays are required to overcome the distance limitations on quantum key establishment. In this case the relays randomly switch on or off as before. However, we now need at least 2 of the relays to be in operation in order to establish a successful quantum key between Alice and Bob. Accordingly, the first 4 channels shown in Table 2 can no longer establish such a key.

**Table 2.** The possible logical channels created using the random drop-out of relays. In this instance at least 2 relays are required to overcome the distance limitation on the channel between Alice and Bob so that a quantum key cannot be established for the first 4 logical channels listed.

| Alice | $R_1$ | $R_2$ | $R_3$ | Bob | Key |
|---|---|---|---|---|---|
| ■ | □ | □ | □ | ■ | — |
| ■ | ■ | □ | □ | ■ | — |
| ■ | □ | ■ | □ | ■ | — |
| ■ | □ | □ | ■ | ■ | — |
| ■ | ■ | ■ | □ | ■ | $K_{12}$ |
| ■ | ■ | □ | ■ | ■ | $K_{13}$ |
| ■ | □ | ■ | ■ | ■ | $K_{23}$ |
| ■ | ■ | ■ | ■ | ■ | $K_{123}$ |

As before Alice and Bob can now consider each of these successful open channels as shares. The final quantum key can therefore be constructed by forming

$$K = K_{12} \oplus K_{13} \oplus K_{23} \oplus K_{123} \quad (2)$$

Again, as before, an attacker would need to compromise all of the relays on the channel in order to obtain the final key. However, because of the distance limitation any single relay on its own never forms part of any key establishment.

As a final example let us consider a channel that requires $N$ relays to overcome the distance limitation. In order to be able to operate a secret sharing scheme we have to build in redundancy. The minimum we can consider is clearly just one extra relay. The final key can therefore be formed from

$$K = K_{123...N+1} \oplus \\ K_{2345...N+1} \oplus K_{1345...N+1} \oplus \\ K_{1245...N+1} \oplus ..... \oplus K_{12345...N} \quad (3)$$

which is just $N + 2$ key shares where $N$ is the number of relays needed to establish the channel. An eavesdropper has to compromise all of the relays on the channel, as before.

So we find the important and useful result that adding a *single extra relay* to a channel which has the minimum number of relays to ensure a successful quantum key exchange guarantees that an eavesdropper has to compromise *all* of the relays on the channel in order to obtain the secret key.

## 3. CONCLUSIONS

We have presented a simple scheme for using secret sharing techniques on a single physical quantum key distribution channel that requires relays for its operation. We find that an attacker has to compromise *all* of the relays on the channel in order to obtain the key. Furthermore, we have shown that *only a single extra relay* need be added to ensure this security. This is an important result that could have a large impact upon the design and operation of quantum key distribution networks.

## 4. REFERENCES

[1] N. Gisin, G. Ribordy, W. Tittel, and H. Zbinden, "Quantum Cryptography", *Rev. Mod. Phys.*, **74**, 145–195, 2002.
[2] S.M. Barnett and S.J.D.Phoenix, "Extending the Reach of QKD Using Relays", *this conference*, 2011.
[3] T.R. Beals and B.C. Sanders, "Distributed relay protocol for probabilistic information theoretic security in a randomly-compromised network", in *Third International Conference on Information Theoretic Security (ICITS) 2008*, *LNCS*, vol 5155, pp. 29–39. Springer, 2008.
[4] A.J. Menezes, P.C. van Oorschot and S.A. Vanstone "Handbook of Applied Cryptography" CRC Press,1996.